\documentclass[doublecol]{epl2} 

\usepackage{psfrag,ulem,csquotes}
\usepackage{amssymb}
\usepackage{amsmath}
\usepackage{color}
\usepackage{hyperref}
\usepackage{mathtools}
\usepackage{color}   
\usepackage{soul}

\def\be{\begin{equation}}
\def\ee{\end{equation}}
\def\bea{\begin{eqnarray}}
\def\eea{\end{eqnarray}}
\def\bml{\begin{mathletters}}
\def\eml{\end{mathletters}}

\title{Polygenic adaptation in changing environments}

\shorttitle{Polygenic adaptation in changing environments} 

\author{Kavita Jain \and Archana Devi}
\shortauthor{Kavita Jain \and Archana Devi}

\institute{Theoretical Sciences Unit, Jawaharlal Nehru Centre for Advanced Scientific Research, Jakkur P.O., Bangalore 560064, India}


\abstract{Although many phenotypic traits are determined by a large number of genetic variants, how  a polygenic trait adapts in response to the changes in the environment is still poorly understood. Here we study the adaptation dynamics of a polygenic trait {that is determined by a finite number of genetic loci} 
in an infinitely large population which is evolving under stabilising selection and recurrent mutations. We find that in a 
changing environment, modeled here by a linearly moving phenotypic optimum, the mean trait also moves linearly with time. But its speed is smaller than that of the phenotypic optimum when the effect sizes of the genetic variants are small and approaches that of the environmental change for larger effect sizes. Our study thus highlights the influence of the genetic architecture of a polygenic trait on its adaptability.}

\begin{document}
\maketitle

\section{Introduction}

Understanding phenotypic variation in terms of the underlying genetic variation is one of the central problems in biological evolution \cite{Barton:2002}.  During the last decade, genome-wide association studies (GWAS) 
 \cite{Visscher:2012} have provided valuable insights into the genetic architecture of phenotypic traits, and the information about the number of genetic variants that affect a phenotype, the size of their effects and their relative frequency is  becoming increasingly available \cite{Timpson:2017}. In some cases such as industrial melanism in peppered moth \cite{Hof:2011}, a  phenotypic trait is determined by one or few genes. But many phenotypic traits ranging from crop yield and human height to complex diseases are polygenic  as they are influenced by a large number of genetic variants. It has been suggested that such quantitative traits may even be omnigenic, determined by all the genes due to the interconnectedness of gene networks {\cite{Boyle:2017,Wray:2018}}. The effect sizes of these genetic variants may be large or small,  and the proportions in which they occur in a trait is governed by evolutionary forces such as selection \cite{Timpson:2017}. 

For several decades, the infinitesimal model \cite{Fisher:1918,Bulmer:1980} has served as  workhorse for describing polygenic adaptation \cite{Kopp:2014}. {In this model, an infinite number of genetic loci each with an infinitesimal  effect underlie a phenotypic trait. More precisely, the effect size is assumed to decrease with increasing number of loci which results} in a genetic variance that remains constant in time in accordance with some phenotypic data [9] and renders the problem analytically tractable. 
{However, the effect sizes can be finite,} 
and genetic variance can change, at least, over some time scales.

Recently a model of polygenic adaptation was introduced \cite{Vladar:2014} in which the phenotypic trait evolves under the action of stabilising selection and mutations. A large {but finite number of loci contribute to the trait, and the effect sizes which may be large or small are different at different loci.} It is important to note that in \cite{Vladar:2014}, an effect size is large or small relative to a scaled mutation rate (defined later) and, unlike in the infinitesimal model, does not depend on the number of loci under selection. 

In this article, we study the adaptation dynamics within the framework of the model in \cite{Vladar:2014} when the phenotypic optimum moves due to a change in the environment. 
The scenario in which the phenotypic optimum shifts suddenly because of, say, a natural disaster was recently studied and an analytical method was developed to take the changes in the genetic variance into account \cite{Jain:2017a,Jain:2017c}; here we consider the case when the phenotypic optimum moves gradually as a result of, for example, climate or societal change \cite{Sanjak:2018}. Adaptation in the face of 
moving phenotypic  optimum has been previously investigated within infinitesimal model ignoring mutations and changes in the genetic variance \cite{Lynch:1993,Burger:1995}. 
Here we find that the adaptation dynamics are strongly affected by mutations {and the number of loci under selection} when the effect sizes are small, and mediated by large transient changes in the genetic variance when the effect sizes are large.

\section{Models}

We study the evolution of a single quantitative trait in an infinitely large population of diploids.  
The phenotypic trait value $z$ is determined by $\ell$ diallelic genetic loci, each of which  contributes to the phenotype in an additive fashion. If the $\pm$ allele at the $i$th locus has an effect $\pm \gamma_i/2$ on the trait, 
the mean phenotypic trait averaged over the population can be written as
\be
c_1=2 \sum_{i=1}^\ell \left(\frac{\gamma_i}{2} \right) p_i+ \left( \frac{-\gamma_i}{2} \right)q_i~,
\label{c1def}
\ee
where $p_i$ and $q_i=1-p_i$, respectively, denote the frequency of $+$ and $-$ allele at the $i$th locus in the population.  In the following, the effect sizes are chosen independently from an exponential distribution with mean ${\bar \gamma}$, as suggested by quantitative-genetic studies \cite{Mackay:2004}. 

We consider the situation when the allele frequencies and phenotypic properties evolve due to stabilising selection and recurrent mutations. The fitness of a phenotypic trait under stabilising selection is maximum at an optimum phenotypic value $z_f$ which, in general, may vary with time (see below);  away from the optimum, the fitness  decreases quadratically, $w(z)=1- (s/2) (z-z_f)^2$ where $0 < s < 1$ is the strength of selection. {Mutations generate variation} and here, the $+$ and $-$ alleles mutate to each other at rate $\mu$. We also assume that recombination occurs faster than selection and therefore the inter-loci correlations can be ignored (linkage equilibrium) \cite{Barton:1991}. Then it can be shown that the allele frequency $p_i$ evolves according to \cite{Vladar:2014}
\be
{\dot p}_i=-s \gamma_i  (c_1-z_f) p_i q_i-\frac{s \gamma_i^2}{2} p_i q_i (q_i-p_i) + \mu (q_i-p_i)~,
\label{maineq}
\ee
where dot denotes the derivative with respect to time. 
In the above equation, the first term on the right-hand side (RHS) that depends on the frequency of {\it all} the loci acts to decrease the deviation between the mean trait and the phenotypic optimum, the second term tends to fix one of the alleles thus depleting genetic variation while the third term compensates for the loss in diversity through mutations.

In the stationary state where ${\dot p}_i=0$, if the deviation between the mean trait and phenotypic optimum is zero, the stable solutions for the equilibrium allele frequencies are given by \cite{Vladar:2014}
\be
{p_i^*=}
\begin{cases}
\frac{1}{2} ~&,~\gamma_i < \hat \gamma \\
\frac{1}{2}\pm\frac{1}{2} \sqrt{1-\frac{{\hat \gamma}^2}{\gamma_i^2}}  ~&,~\gamma_i > \hat \gamma~,
\end{cases}
\label{equil}
\ee
where ${\hat \gamma}=\sqrt{8 \mu/s}$. Thus when the effect size is smaller than the threshold effect ${\hat \gamma}$, both the alleles are present in the population whereas for larger effect sizes, one of the alleles is close to fixation. {The steady state deviation in the mean trait from its optimum need not be zero and can be estimated by approximating the effect size of all the loci by the average effect ${\bar \gamma}$. From (\ref{maineq}) in the steady state, we then find the mean equilibrium trait  to be $c_1^*/z_f \approx \left[1 + 4 \mu (s \ell {\bar \gamma}^2)^{-1} \right]^{-1}$.}

In this article, we are interested in the dynamics of the mean trait when the phenotypic optimum moves to a new value. From (\ref{maineq}) for the allele frequency, it can be shown that the time evolution equation  for the mean trait depends on the variance and skewness of the phenotypic trait and, in general, the dynamics of a trait cumulant depend on two higher cumulants \cite{Burger:2000}. This cumulant hierarchy makes it difficult to obtain analytical results for the dynamics. 
However, recent work \cite{Jain:2015,Jain:2017a} has established that the bulk of the adaptation process is driven by  
 the deviation between the mean trait and the phenotypic optimum. Thus, at short times, it is a good approximation to  
 retain only the first term on the RHS of (\ref{maineq}); this yields the {\it directional selection model} in which \cite{Jain:2017a}
\be
{\dot p}_i \approx -s \gamma_i (c_1-z_f) p_i q_i ~,~i=1,..., \ell~.
\label{dseq}
\ee

The above equation shows that the mean trait evolves as 
\be
{\dot c_1}= -s c_2 (c_1-z_f)~,
\label{c2eqn}
\ee
where the genetic variance $c_2= 2\sum_{i=1}^\ell \gamma_i^2 p_i q_i$. If the genetic variance in (\ref{c2eqn}) remains constant in time, we arrive at the Lande's equation for the evolution of the mean trait \cite{Lande:1983a} which is obviously solvable. But even for time-dependent $c_2$ and other trait cumulants, the directional selection model is analytically tractable as briefly described below \cite{Jain:2017a}. We first note that for any two loci $k$ and $j$,
\be
\frac{dp_j}{\gamma_j p_j q_j}=\frac{dp_k}{\gamma_k p_k q_k}=-s (c_1-z_f) dt~.
\ee
From the first equality,  we find that the allele frequency at the $j$th locus can be expressed in terms of its initial frequency and that at the $k$th locus, 
 \be
p_j(t)=1-\frac{1}{1+ \frac{p_j(0)}{q_j(0)} \left( \frac{p_k(t) q_k(0)}{p_k(0) q_k(t)} \right)^{\gamma_j/\gamma_k}}~.
\ee
Using this in (\ref{c1def}), one can write the mean phenotypic trait as
\be
c_1=\sum_{i=1}^\ell \gamma_i - \sum_{i=1}^\ell  \frac{2 \gamma_i}{1+ \frac{p_i(0)}{q_i(0)} e^{\beta \gamma_i/{\bar \gamma}}}~,
\label{c1ds}
\ee
where 
\be
\beta(t)= \frac{\bar \gamma}{\gamma_k} \ln \left( \frac{p_k(t) q_k(0)}{p_k(0) q_k(t)}\right)
\label{betadef}
\ee
is locus-independent and, due to (\ref{dseq}), evolves as
\be
{\dot \beta}= -s {\bar \gamma} (c_1-z_f)~.
\label{betaeqn}
\ee
Equations (\ref{c1ds}) and (\ref{betaeqn}) together yield a closed equation for the mean trait. 

\section{Results}

We start with a population equilibrated to the phenotypic optimum at $z_0$. {The phenotypic optimum moves linearly  until a time $\tau$ when it reaches $z_\tau > z_0$ where it stays for later times. Thus the new phenotypic optimum $z_f(t)=z_0+ v t, t < \tau$ where the velocity $v=(z_\tau-z_0)/\tau$. Here, $|z_0|, |z_{\tau}| < {\ell {\bar \gamma}}$ as the mean trait is bounded for finite $\ell$. 
Our goal is to understand how the population adapts while the environment is changing. In the following, all the numerical data are obtained for a single realisation of effects as the quantities of interest are expected to be self-averaging \cite{Jain:2017a}.}

\subsection{When most effects are small (${\bar \gamma} \ll  {\hat \gamma}$)}

\begin{figure}
\onefigure[width=85mm]{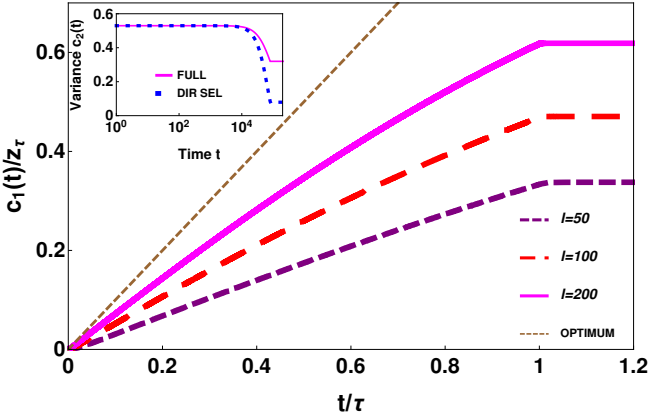}
\caption{{Evolution of mean trait (main) in the full model when most effects are small (${\bar \gamma}=0.05 \ll  {\hat \gamma} \approx 0.89$) and the phenotypic optimum moves with a constant speed $v=10^{-4}$ for various $\ell$. Here $s=10^{-2}, \mu=10^{-3}, z_0=0$ and $z_\tau=0.8 \ell {\bar \gamma}$. {The phenotypic optimum $z_f(t)/z_\tau$ is also shown for comparison.} 
The inset shows the evolution of genetic variance for $\ell=200$.}}
\label{fig1}
\end{figure}

Figure~\ref{fig1} shows that for linearly moving phenotypic optimum, the mean trait also increases linearly with time at a speed that decreases as the number of loci determining a trait decrease. However, the genetic variance $c_2$ (shown in inset) remains close to its stationary value $c_2^* \approx \ell {\bar \gamma}^2$ \cite{Vladar:2014} until time $\tau$ and then decreases to the stationary genetic variance {corresponding to the equilibrium mean trait $c_1^*$ \cite{Jain:2017a}. For the parameters in Fig.~\ref{fig1}, we expect $c_1^*/z_\tau$ to be $0.24, 0.39, 0.56$ (see the discussion following (\ref{equil})) which is close to $c_1(\tau)/z_\tau=0.33, 0.47, 0.62$ for $\ell = 50, 100$ and  $200$, respectively. For this reason, the mean trait does not change substantially when the phenotypic optimum stops moving.}

For $t \ll \tau$ where the genetic variance is approximately constant, (\ref{c2eqn}) for the evolution of mean trait simplifies to ${\dot c_1}(t) \approx s c_2^* (z_f(t)-c_1(t))$  which can be readily integrated to yield 
\be
c_1(t)=z_0+ v t - \frac{v}{s c_2^*} (1-e^{-s c_2^* t})~,~t <  \tau~,
\label{lagsm}
\ee
and shows that the mean trait always lags behind the moving phenotypic optimum.  At short times ($\ll (s c_2^*)^{-1}$), the mean trait increases quadratically. 
But at longer times, $c_1$ increases linearly with speed $v$ and the lag $z_f(t)-c_1(v,t)$ reaches a {constant} value $v/(s c_2^*)$. This result has also been obtained  in infinitesimal model \cite{Lynch:1993,Burger:1995} in which an infinite number of loci, each with an infinitesimal effect ($\sim \ell^{-1/2}$), contribute to a phenotypic trait 
{and the mean trait keeps increasing in an unbounded fashion; here, as a finite number of loci contribute to the trait, the linear behaviour of the mean trait sets in at a time $\sim \ell^{-1}$ and continues until the time $\tau$.}

\begin{figure}
\onefigure[width=85mm]{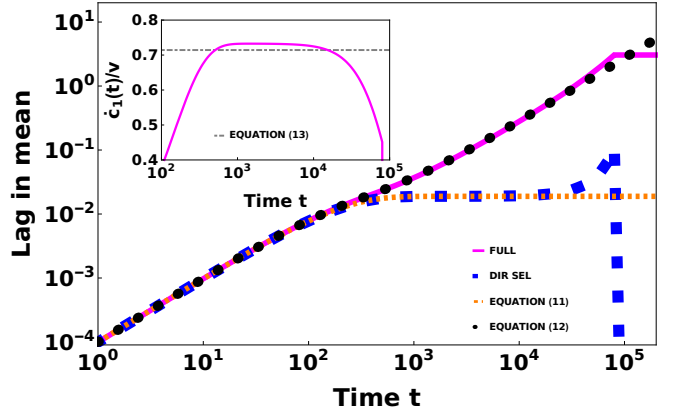}
\caption{Lag in mean when most effects are small (main). While the lag is constant in the directional selection model, it increases in the full model. This is because the mean trait moves slower than the phenotypic optimum (inset). Here $\ell=200$ and rest of the parameters are the same as in Fig.~\ref{fig1}.}
\label{fig2}
\end{figure}

Figure~\ref{fig2} shows that for $(s c_2^*)^{-1} \ll t \ll \tau$, the lag is constant in the directional selection model and matches the prediction (\ref{lagsm}) but it keeps increasing linearly with time in the full model. To understand this behavior, we note that at short times, by virtue of (\ref{equil}), the allele frequencies are close to one half and therefore the last two terms on the RHS of (\ref{maineq}) can be ignored \cite{Jain:2017a}. Indeed, as Fig.~\ref{fig2} shows, the full model and the directional model are in good agreement at short times. But at longer times, {the allele frequencies are not in equilibrium and for $z_\tau > z_0$, as is assumed here, the frequency of the $+$ allele increases towards one.} Moreover, when the effects are small, the mutation rate is large ($\mu > s \gamma_i^2/8$). These considerations suggest to modify the directional selection model (\ref{dseq}) by adding the mutation term to it when the phenotypic optimum is moving. The mean trait then evolves according to 
\be
{\dot c_1}(t) \approx s c_2^* (z_f(t)-c_1(t))-2 \mu c_1(t)~,
\label{modDS}
\ee
which shows that at large times, the mean trait increases linearly as $c_1(t)= u t+ \lambda {~,~ (s c_2^*+ 2 \mu)^{-1} < t < \tau}$, where
\bea
u &=& \frac{v}{1+ 2 \mu (s c_2^*)^{-1}}=\frac{v}{1+ \frac{1}{\ell} \left(\frac{\hat{\gamma}}{2 {\bar \gamma}} \right)^2} \label{speedu} \\
\lambda &=& \frac{z_0-u  (s c_2^*)^{-1}}{1+ 2 \mu (s c_2^*)^{-1}}~. \label{lagu}
\eea 
The inset of Fig.~\ref{fig2} shows that the rate of change of mean trait is in agreement with the speed $u$. 

{Equation (\ref{speedu}) above makes two important points. First, the mean trait in the full model (\ref{maineq}) increases {\it slower} than the phenotypic optimum with increasing mutation rates. At short times where the allele frequencies are close to their equilibrium value given by one half, mutations occurring at equal rate between the $+$ and $-$ allele do not affect the dynamics. But at larger times, as the $+$ alleles become more abundant than the $-$ alleles owing to selection, the net effect of the mutations is deleterious leading to an increased lag in the mean trait. Second, the speed of the mean trait increases with the number of loci under selection. When ${\hat \gamma}, {\bar \gamma}$ are of order one (as is the case here), the genetic variance is much larger than the mutations; this holds in the infinitesimal model also where the mean effect decreases with $\ell$ keeping $\mu \ell$ finite. In either case, 
 mutation is weaker than selection and may be neglected when $\ell \to \infty$. But for finite number of loci, mutations also enter the picture and, as argued above, decrease the speed of the mean trait.}

\subsection{When most effects are large (${\bar \gamma} \gg  {\hat \gamma}$)}

\begin{figure}
\onefigure[width=85mm]{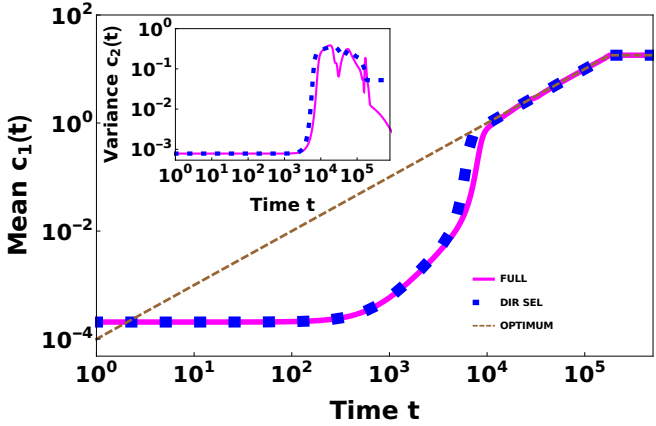}
\caption{Evolution of mean (main) and genetic variance (inset) when most effects are large (${\bar \gamma}=0.1 \gg {\hat \gamma} \approx 0.0028$) and the phenotypic optimum moves with a constant speed $v=10^{-4}$. {Here $\ell=200, s=10^{-2}, \mu=10^{-8}, z_0=0$ and $z_\tau=18$}.}
\label{fig3}
\end{figure}
\begin{figure}
\onefigure[width=85mm]{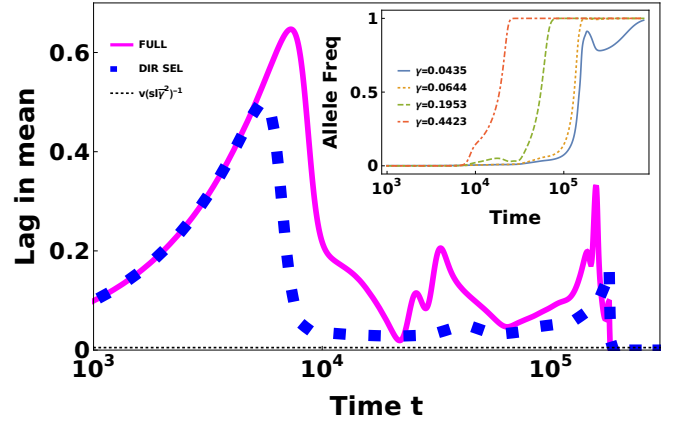}
\caption{Lag in mean (main) and allele frequency dynamics (inset) when most effects are large. All the parameters are the same as in Fig.~\ref{fig3}.}
\label{fig4}
\end{figure}

Since the mutations are unimportant when effect sizes are large, the directional selection model (\ref{dseq}) describes the full model well (see Figs.~\ref{fig3} and \ref{fig4}). The inset of Fig.~\ref{fig3} shows that the genetic variance remains close to its equilibrium value $c_2^*=\ell {\hat \gamma}^2$ \cite{Vladar:2014} for some time, rises quickly to ${\tilde c_2} \sim 100 c_2^*$ where it stays before finally dropping to $c_2^*$. As $c_2^*$ is very small for the parameters in Fig.~\ref{fig3}, due to (\ref{c2eqn}), the mean trait remains close to $z_0$ initially and then increases before {saturating to $c_1^* \approx z_\tau$.} Since the genetic variance changes during the time intervals of interest,  (\ref{c2eqn}) for  the mean trait evolution does not close and 
therefore we now work with (\ref{c1ds}) and (\ref{betaeqn}).

As the equilibria are bistable for large effects (see (\ref{equil})), a fraction $f$ of the initial population carries $+$ allele which is related to the initial phenotypic mean through $z_0 \approx (2 f-1) \ell {\bar \gamma}$. We may therefore write the mean trait  (\ref{c1ds}) as
\bea
{ c_1}=\sum_{i=1}^\ell \left(\gamma_i -  \frac{2 \gamma_i f}{1+ \frac{p_{i,+}^*}{q_{i,+}^*} e^{{ \beta}\gamma_i/{\bar \gamma}}} -\frac{2 \gamma_i (1-f)}{1+ \frac{p_{i,-}^*}{q_{i,-}^*} e^{{ \beta} \gamma_i/{\bar \gamma}}}\right)~,
\label{c1le1}
\eea
which, for large $\ell$, simplifies to 
\bea
\frac{{ c_1}(t)}{\ell {\bar \gamma}} \approx 1- \left(1+\frac{z_0}{\ell {\bar \gamma}} \right) I_{+}({ \beta})
- \left(1-\frac{z_0}{\ell {\bar \gamma}} \right) I_{-}({ \beta})~,
\label{z0eff}
\eea
where
\be
I_{\pm}({ \beta})=\int_{2/\alpha}^\infty dx \frac{x e^{-x}}{1+ (\alpha x)^{\pm 2} e^{{ \beta} x}}
\ee
and $\alpha=2 {\bar \gamma}/{\hat \gamma} \gg 1$. Equation (\ref{z0eff}) generalises equation 24 of \cite{Jain:2017a} where the initial phenotypic optimum is taken to be zero. 

We first note that ${\dot \beta}, \beta \geq 0$ - the first assertion follows from the intuitive expectation that the lag must always be nonnegative and (\ref{betaeqn}), and the second one on using $\beta(0)=0$ (see (\ref{betadef})). 
Then it is easy to see that for large $\alpha$, the integral $I_+ \to 0, I_- \to 1$. Neglecting $I_+$ in (\ref{z0eff}) and writing $I_-=\int_{2/\alpha}^\infty dx~ x e^{-x}-J_- \approx 1- J_-$, we  find that the parameter $\beta$ evolves according to 
\be
{\dot { \beta}} \approx s {\bar \gamma} (\ell \bar \gamma -z_0) \left[ \rho(t)  - J_-({ \beta})\right]~,
\label{betaapp}
\ee
where
\be
J_-=\int_{2/\alpha}^\infty dx \frac{x e^{-x}}{1+ (\alpha x)^{2} e^{-{ \beta} x}}
\ee 
and $\rho(t)= {v t}/{(\ell {\bar \gamma}-z_0)}$. For $\beta \ll 1$, the integral $J_- \approx 0$ as it is heavily suppressed by the factor $\alpha^{2}$ in the denominator of the integrand \cite{Jain:2017a}. 
For $\beta \gg 1$, the integral $J_-$ may be estimated by approximating the factor $(1+ (\alpha x)^{2} e^{-{ \beta} x})^{-1}$ in the integrand of $J_-$ by a Heaviside theta function $\Theta(x-{x_-})$  where $x_-(\beta)$ is a solution of $\alpha^2 {x^2_-} e^{-{ \beta} x_-}=1$.  We thus obtain $J_-\approx (1+{x_-}) e^{-x_-}~,~\beta \gg 1$ \cite{Jain:2017a}. 

We are now in a position to understand the behaviour of the mean trait given by 
\be 
c_1(t)-z_0=(\ell {\bar \gamma}-z_0) J_-(\beta)=v t - \frac{{\dot \beta}}{s {\bar \gamma}}~,
\label{c11}
\ee
where we have used (\ref{betaapp}) to express $J_-$ in terms of ${\dot \beta}$. 
At short times ($t < \tau_0$), as the mean stays close to its initial value $z_0$,  the integral $J_- \approx 0$ and therefore  
$\beta(t)=(t/\tau_0)^2$ where $\tau_0=\sqrt{2/(s v {\bar \gamma})}$. After time $\tau_0$, there is a large increase in the genetic variance and the mean trait starts evolving. {Note that if the genetic variance was held constant at its initial value, as in the infinitesimal model, the mean trait will stay at $z_0$ and the population can not adapt.}

In the time interval {$(s {\tilde c_2})^{-1} < t < \tau$} where the mean trait increases linearly, the genetic variance ${\tilde c_2} \gg c_2^*$ remains roughly constant. To estimate ${\tilde c_2}$, we note from the inset of Fig.~\ref{fig4} 
that the allele frequencies at loci with effect size larger than ${\bar \gamma}$ sweep to fixation much earlier than  at loci with smaller effect sizes. {For the set of effects used in Fig.~\ref{fig4}, the allele frequencies at $3$ loci with effect size $\approx 0.4$ swept to fixation while nearly $30$ selective sweeps occurred when the optimum was shifted suddenly to $z_\tau$ (data not shown). Large changes in allele frequency at many loci  in the latter case is consistent with the fact that the population adapts exponentially fast over a short time scale (that decreases with $\ell$) when the optimum shifts suddenly \cite{Jain:2017a}; here the mean deviation changes slowly over a time of order $\ell$ and besides a few sweeps, many small frequency shifts ($\sim 0.1$) also occur. While the selective sweep is in progress at loci  with very large effects,  the genetic variance rises to ${\tilde c}_2 \sim \ell \int_{\bar \gamma}^\infty d\gamma \gamma^2 p(\gamma) \sim \ell {\bar \gamma}^2$. Then replacing $c_2^*$ by ${\tilde c_2}$ in (\ref{speedu}) and using that ${\hat \gamma} \ll {\bar \gamma}$ when effects are large, it follows that, as in the infinitesimal model, the mean trait increases with speed $v$ and lag $v/(s {\tilde c_2})$.}

We now perform a more careful analysis of the lag in mean. The above discussion shows that the lag ${\dot \beta}/{s {\bar \gamma}}$ (second equality in (\ref{c11})) is small for $\ell \gg 1$.  
Therefore, to a first approximation, we may write $\rho(t) \approx J_-(\beta)$ in (\ref{betaapp}) which yields 
\be
{x_-}(t)=-1-W_{-1}\left(-\frac{\rho(t)}{e} \right)~,
\label{xminus}
\ee
where $W_{-1}(x)$ is the lower branch of the Lambert W function \cite{Corless:1996}. 
Using $\beta=(2/x_-) \ln (\alpha x_-)$, we find  that 
\be
{\dot \beta}=\frac{2}{t} \frac{W_{-1}(-\rho/e)}{[1+W_{-1}(-\rho/e)]^3} \ln \left(\frac{1+W_{-1}(-\rho/e)}{-e/\alpha} \right)~. 
\ee
For $t \ll \tau$ where $\rho/e \ll 1$, the function $W_{-1}(-\rho/e) \approx \ln(\rho/e) \sim \ln t$ \cite{Corless:1996} so that ${\dot \beta} \sim t^{-1}$. Thus when most effects are large, the mean trait moves with speed $v$ but the lag between the mean trait and phenotypic optimum is not a constant.  Although the lag $\sim (s {\bar \gamma} t)^{-1}$ decays rapidly with time, it is larger than that obtained within the infinitesimal model for a polymorphic population since the above discussion is valid for $t < \tau$, see Fig.~\ref{fig4}. {Furthermore,  (\ref{c2eqn}) and (\ref{betaeqn}) show that the genetic variance $c_2(t) =(s v {\bar \gamma}-{\ddot \beta})/(s^2 {\bar \gamma} z_0+ s{\dot \beta})$ which on using the above results for $\beta$ yields $c_2(t) \sim v {\bar \gamma} t$. Thus on time scales of order $\tau$, we obtain the genetic variance to be ${\tilde c_2}$ in agreement with the argument above. Since the genetic variance increases from $c_2^*$ to ${\tilde c}_2$, for fixed ${\hat \gamma}, {\bar \gamma}$, a large change in variance occurs with increasing $\ell$. But the change in variance is smaller if the mean effect and the threshold effect are not substantially different (see Fig. 5 below).} 


\section{Summary and open questions}

\begin{figure}
\onefigure[width=85mm]{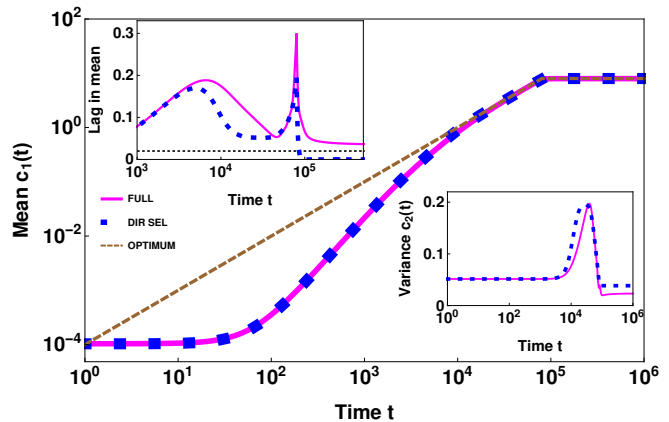}
\caption{Evolution of mean (main), lag in mean (top left inset) and genetic variance (bottom right inset) when equal number of effects are small and large (${\bar \gamma}=0.05, {\hat \gamma} \approx 0.028$) and the phenotypic optimum moves with a constant speed $v=10^{-4}$. {Here $\ell=200, s=10^{-2}, \mu=10^{-6}, z_0=0$ and $z_\tau=8$.} The dotted line in the inset is the minimum lag $v (s \ell {\bar \gamma}^2)^{-1}$.}
\label{fig5}
\end{figure}


While stabilising selection has the tendency to pull the population towards the phenotypic optimum, recurrent mutations create variants that push the population away from it.  
{The behaviour of the mean trait in the large-effects case where selection is the dominant process resembles that in the infinitesimal model as the mean trait moves with the speed of the phenotypic optimum, while in the small-effects case where mutation rates are large, the mean trait moves slower than the phenotypic optimum. However, in the latter case, as the number of genetic loci affecting a trait increases, selection dominates over mutations and the speed of the mean trait approaches that of the phenotypic optimum  (also, see the discussion after (\ref{lagu}))}. 

In the large-effects case as the initial variance is small, the mean trait {stays close to its initial value  resulting in large deviation from the moving phenotypic optimum. But this causes the allele frequencies to change in a manner that increases the genetic variance thus paving the way for adaptation.} 
On the other hand, when most effects are small, there is sufficient initial genetic variance for adaptation to proceed keeping genetic variance constant. 

In the last section, we studied two limiting cases of the genetic architecture, {\it viz.}, when ${\bar \gamma}/{\hat \gamma}$ is much smaller or larger than unity.  Figure~\ref{fig5} shows that when both small and large effect sizes constitute a phenotypic trait, the behaviour of the mean trait is closer to the mostly large-effects case, possibly because the contribution of the small effect sizes to the mean trait increases slower than that of the large-effects loci. 
Thus we conclude that, in general, the lag between mean trait and phenotypic optimum exceeds that predicted by the infinitesimal model for a polymorphic population \cite{Lynch:1993,Burger:1995}.

Although the study presented here goes beyond the standard quantitative-genetic theory \cite{Lynch:1993,Burger:1995} by accounting for mutations and temporal changes in genetic variance,  this work also has some limitations - we have neglected the effect of epistasis in the genotype-phenotype map, pleiotropic effects of other phenotypic traits and, perhaps most importantly, the finiteness of the population size. The effect of stochastic fluctuations on the speed of adaptation and extinction risk has been addressed in recent work 
\cite{Jones:2004,Matsuzewski:2015} when the genetic variance remains constant. Extending the results presented here to finite populations is desirable and we plan to address this in a future work. 

\acknowledgments {We thank an anonymous reviewer for useful suggestions and Wolfgang Stephan for comments.}



\begin{thebibliography}{10}
\expandafter\ifx\csname url\endcsname\relax\def\url#1{\texttt{#1}}\fi

\bibitem{Barton:2002}
\Name{Barton N.~H. \and Keightley P.~D.} \REVIEW{Nat. Rev.
  Genet.}{3}{2002}{11}.

\bibitem{Visscher:2012}
\Name{Visscher P.~M., Brown M.~A., McCarthy M.~I. \and Yang J.} \REVIEW{Am J
  Hum Genet.}{90}{2012}{7}.

\bibitem{Timpson:2017}
\Name{Timpson N.~J., Greenwood C. M.~T., Soranzo N., Lawson D.~J. \and Richards
  J.~B.} \REVIEW{Nat. Rev. Genet.}{19}{2018}{110}.

\bibitem{Hof:2011}
\Name{van't Hof A.~E., Edmonds N., Dalikov{\'a} M., Marec F. \and Saccheri
  I.~J.} \REVIEW{Science}{332}{2011}{958}.

\bibitem{Boyle:2017}
\Name{Boyle E.~A., Li Y.~I. \and Pritchard J.~K.}
  \REVIEW{Cell}{169}{2017}{1177}.

\bibitem{Wray:2018}
\Name{Wray N.~R., Wijmenga C., Sullivan P.~F., Yang J. \and Visscher P.~M.}
  \REVIEW{Cell}{173}{2018}{1573}.

\bibitem{Fisher:1918}
\Name{Fisher R.~A.} \REVIEW{Trans. Roy. Soc. Edinburgh}{52}{1918}{399}.

\bibitem{Bulmer:1980}
\Name{Bulmer M.~G.} \Book{The Mathematical Theory of Quantitative Genetics}
  (Oxford University Press, Oxford) 1980.

\bibitem{Kopp:2014}
\Name{Kopp M. \and Matuszewski S.} \REVIEW{Evolutionary
  Applications}{7}{2014}{169}.

\bibitem{Lande:1976}
\Name{Lande R.} \REVIEW{Evolution}{30}{1976}{314}.

\bibitem{Vladar:2014}
\Name{de~Vladar H.~P. \and Barton N.~H.} \REVIEW{Genetics}{197}{2014}{749}.

\bibitem{Jain:2017a}
\Name{Jain K. \and Stephan W.} \REVIEW{Genetics}{206}{2017}{389}.

\bibitem{Jain:2017c}
\Name{Jain K. \and Stephan W.} \REVIEW{Mol. Biol. Evol}{34}{2017}{3169}.

\bibitem{Sanjak:2018}
\Name{Sanjaka J.~S., Sidorenkoc J., Robinson M.~R., Thornton K.~R. \and
  Visscher P.~M.} \REVIEW{Proc. Natl. Acad. Sci. USA}{115}{2018}{151}.

\bibitem{Lynch:1993}
\Name{Lynch M. \and Lande R.} in \Book{Biotic Interactions and Global Change},
  edited by \Name{Kareiva P.~G., Kingsolver J. \and Huey R.~B.} (Sinauer
  Associates, Sunderland, MA) 1993 pp. 234--250.

\bibitem{Burger:1995}
\Name{B{\"u}rger R. \and Lynch M.} \REVIEW{Evolution}{49}{1995}{151}.

\bibitem{Mackay:2004}
\Name{Mackay T. F.~C.} \REVIEW{Current Opinion in Genetics and
  Development}{14}{2004}{253}.

\bibitem{Barton:1991}
\Name{Barton N.~H. \and Turelli M.} \REVIEW{Genetics}{127}{1991}{229}.

\bibitem{Burger:2000}
\Name{B{\"u}rger R.} \Book{The Mathematical Theory of Selection, Recombination,
  and Mutation} (Wiley, Chichester) 2000.

\bibitem{Jain:2015}
\Name{Jain K. \and Stephan W.} \REVIEW{G3: Genes, Genomes,
  Genetics}{5}{2015}{1065}.

\bibitem{Lande:1983a}
\Name{Lande R.} \REVIEW{Heredity}{50}{1983}{47}.

\bibitem{Corless:1996}
\Name{Corless R., Gonnet G., Hare D., Jeffrey D. \and Knuth D.}
  \REVIEW{Advances in Computational Mathematics}{5}{1996}{329}.

\bibitem{Jones:2004}
\Name{Jones A.~G., Arnold S.~J. \and B{\"u}rger R.}
  \REVIEW{Evolution}{58}{2004}{1639}.

\bibitem{Matsuzewski:2015}
\Name{Matuszewski S., Hermisson J. \and Kopp M.}
  \REVIEW{Genetics}{200}{2015}{1255}.

\end{thebibliography}
\end{document}